# Cathodes and Shape Modification of Cavity for DESY Superconducting Photoinjector*

J. Sekutowicz**, D. Bazyl, E. Vogel, D. Reschke, A. Brinkmann, D. Klinke, D. Kostin, T. Ramm, A. Sulimov, H. Weise, M. Wiencek

*Deutsches Elektronen-Synchrotron DESY, Notkestraße 85, D-22607 Hamburg, Germany*
(Dated: July 2, 2025)

Four DESY prototypes of the L-band superconducting RF (SRF) photoinjector cavity demonstrated on-axis peak gradients ≥ 55 MV/m during multiple vertical cryogenic tests. Two of these prototypes—16G09 and 16G10—achieved these gradients with both superconducting and normal-conducting metallic cathodes, fabricated from either high-purity (RRR = 300) niobium or lower-purity (RRR = 30) copper. The DESY photoinjector, under development for over two decades as a continuous-wave (cw) electron source for FELs, differs from other SRF injectors in that its metallic cathode plug is attached directly to the cavity backplate, exposing the emitting surface to the high electric field within the cavity. This design obviates the need for a choke filter or load-lock system, as presented in 2005 [1]. The initial 1.6-cell cavity geometry was derived from the Low-Loss design developed for the CEBAF 12 GeV upgrade [2], scaled from 1.5 GHz to 1.3 GHz. This shape was later replaced with the current High Gradient TESLA profile. In this report, we discuss current cathode options and present modifications to the cavity shape aimed at significantly reducing the electric field near the cathode opening.

## I. INTRODUCTION

Figure 1 shows CAD cross-section of the 1.6-cell DESY gun cavity with attached cathode plug. Since 2011, we have built ten prototypes of this cavity. Two additional units are currently being manufactured and are expected to be ready for testing later this year. All twelve prototypes are made from high-purity niobium (Nb) with an RRR of 300. Following fabrication, the cavities undergo field flatness adjustment for the π-mode and final BCP treatment. The Nb cathode plug is then attached, after which High-Pressure Rinsing (HPR) is performed to remove particulates from the internal surface of the assembly. This reduces the likelihood of field emission and multipacting. Subsequently, each cavity undergoes vertical testing (VT) at 2 K. If the test results meet the performance criteria, the cavity is approved for further testing. If not, the surface preparation procedure is repeated. Figure 2 shows representative accepted VT results: quality factor ($Q_0$) as a function of $E_p$ (the peak electric field on the axis of the 0.6-cell). Until 2022, high-performing gun prototypes were used to test Nb cathode plugs coated with lead (Pb) on the emitting surface. Lead is a Type I superconductor with a critical temperature of 7.2 K. The highest quantum efficiency (QE) achieved for arc-deposited lead layers was measured at BNL in 2005, yielding QE values of $5 \cdot 10^{-4}$@257 nm and $5 \cdot 10^{-3}$@193 nm [1, 3]. These values exceeded the QE of polycrystalline niobium, as measured and published by the BNL/AES/TJNAF teams in 2003 and 2005 [4, 5, 6]. Despite extensive testing of Pb-coated Nb plugs, the lead coating exhibited poor adhesion to the polished niobium surface. None of the applied deposition methods resulted in a mechanically stable lead layer. Furthermore, poor adhesion precluded the use of HPR as the final surface preparation step. These issues ultimately compelled us to seek an alternative to Pb-coated niobium plugs.

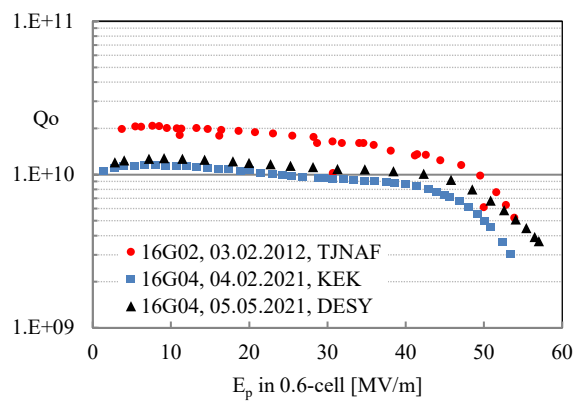

Figure 2: Vertical test results for π-mode of two DESY gun prototypes 16G02 and 16G04, measured at TJNAF, KEK and DESY. The tests were performed at 2K with attached Nb-plugs.

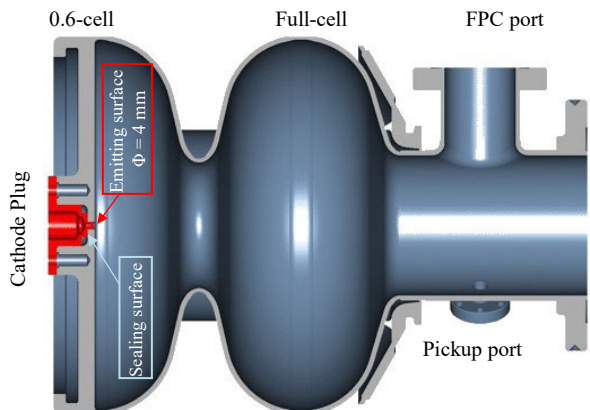

Figure 1: DESY 1.6-cell photoinjector cavity with a cathode plug (red) attached to the backplate of 0.6-cell.

*Work performed in the framework of R&D for future High Duty Cycle operation modes of the European XFEL accelerator, supported by the European XFEL GmbH.
**Corresponding author: jacek.sekutowicz@desy.de

## II. Alternative Cathodes

In 2022 we initiated studies for cathodes made of bulk copper, enabling similarly to niobium HPR for cavity with attached cathode plug. Bulk lead and magnesium were not considered, as both metals are soft and would deteriorate under the 100-bar water jets used in HPR. Their erosion would compromise the cleanliness of the cavity surface, degrading its intrinsic quality factor Qo and gradient.

### a. Cathode-plug made of copper

Copper photocathodes have been extensively studied and implemented in photoinjectors for over three decades. Foundational work is reported in [7, 8, 9]. For nearly 18 years, since its commissioning, the LCLS at SLAC has operated using polycrystalline Cu cathodes as its electron source [10]. The current cathode routinely sustains gradients exceeding 100 MV/m, delivering electron bunches of up to 250 pC at a repetition rate of 120 Hz.

Our thermal modelling in 2022 for the DESY gun cavity with an attached copper plug indicated that a normal-conducting cathode would not impair the SRF performance of the gun cavity. This modelling was validated experimentally in 2023 through a series of vertical tests on the 16G09 and 16G10 prototypes, as reported in [11]. Figures 3 and 4 present examples of these vertical tests conducted at 2 K. These results will be discussed in further detail in Section II.

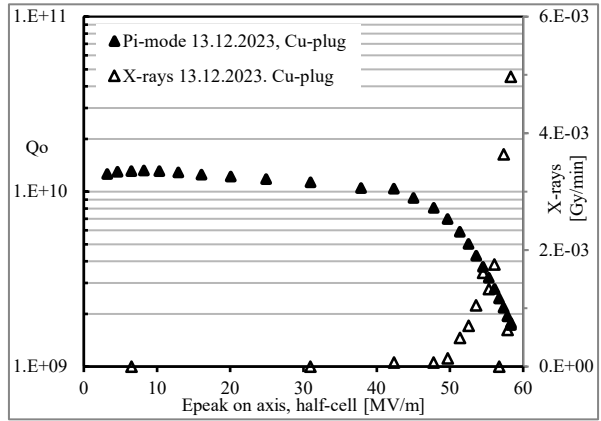

Figure 3: Vertical tests at 2 K of the 16G09 prototype with attached Cu-plug made from low RRR=30 copper.

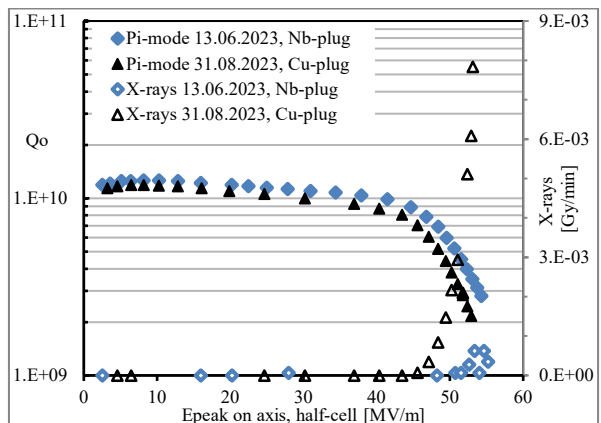

Figure 4: Vertical tests at 2 K of the 16G10 prototype with attached Nb-plug made from high RRR=300 niobium and Cu-plug made from low RRR=30 copper.

Copper exhibits significantly higher thermal conductivity than niobium at 2 K, and this conductivity improves with increasing purity and RRR value [12]. Although a normal-conducting copper cathode dissipates more RF power than a superconducting niobium one, the temperature rise is minimal. This is attributed to the low magnetic field at the cathode surface, copper's high thermal conductivity, and the inclusion of thermal conduction channels in the plug, which facilitate efficient heat transfer to the liquid helium bath. Figure 5 shows a CAD model of the cathode plug used for both Cu and Nb cathodes.

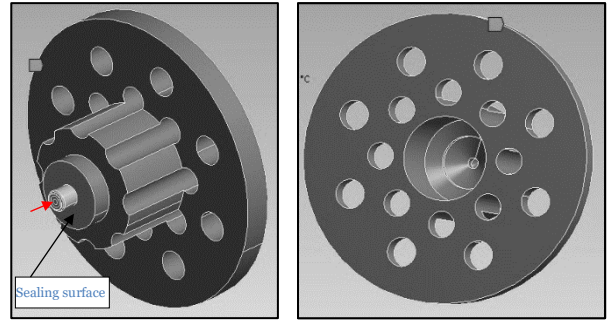

Figure 5: CAD drawing of the cathode-plug. The red arrow indicates the circular surface emitting electrons, which will be irradiated with a UV laser. The tinny cathode cylinder (marked with lighter grey colour), has OD 4 mm and length ca. 2.5 mm. The entire cathode surface exposed to H field is only 44 mm$^2$.

To generate an electron beam with an energy $E_b \geq 4$ MeV $E_p$ on axis in both cells must reach at least 40 MV/m. For the current geometry of the cavity, the electric field at the cathode's emitting surface is $E_c = 35$ MV/m. $E_c$ is lower than $E_p$, due to retraction of the cathode into its channel. Figure 6 shows surface temperature for cathode made from RRR=100 copper. According to the model, when the gun operates at $E_p = 40$ MV/m, the cathode dissipates 60 mW, causing its temperature to rise from 2 K to 2.3 K.

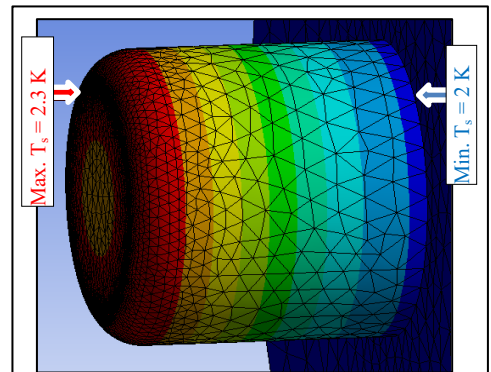

Figure 6: Surface temperature of cathode plug made from RRR=100 copper. The RF-power dissipation in the surface was calculated at $E_p = 35$ MV/m for surface resistance of Rs = 1.04 m$\Omega$ at 1.3 GHz. The plug is cooled with LHe at 2 K.

The quantum efficiency of metallic cathodes is generally much lower than that of alkali and semiconductor cathodes. For metals, QE is primarily influenced by the work function (WF), the wavelength λ of the incident laser, the electric field at the cathode surface $E_c$ (due to the Schottky effect), and the crystal orientation in the case of monocrystalline metals, as discussed in [9]. QE can also vary with temperature and is often highly sensitive to the surface's roughness, oxidation, and contamination.

Table 1 lists the work function and QE of copper for various crystal orientations (i, j, k), evaluated at ambient temperature and 2 K, when irradiated with 4.82 eV photons (λ = 257 nm). QE was calculated using the formula proposed by Dowell and Schmerge [13], along with optical data from [14–17]. For each temperature, we used an effective work function $WF_{eff}$, WF at $E_c = 0$ reduced due to the Schottky effect. This reduction corresponds to $E_c = 1.5$ MV/m during cathode cleaning at room temperature, and $E_c = 35$ MV/m during SRF gun operation at 2 K. However, the QE values at 2 K are theoretical and must be validated experimentally, because the two experiments discussed below lead to contradictory conclusions.

Figure 7 shows the photon absorptivity as a function of wavelength λ for polycrystalline copper and α-Brasses. The data, measured at 4.2 K and presented by Biondi in [18, 19, 20], indicate that for copper (Curve A), the absorptivity at λ ≈ 2600 Å (260 nm) is approximately 63%. Moreover, the study reported in [20] demonstrated no significant difference in photon absorptivity between single-crystal and polycrystalline copper.

Contrary to these findings, the authors of [21, 22] presented experimental data on the QE as a function of temperature for polycrystalline copper and niobium. The investigated temperature range was from 85 K to 400 K. Both metals were irradiated using a laser with λ = 266 nm (4.66 eV photons). Figure 8 shows the fitted curve proposed by the authors for the measured QE of copper, which suggests that QE at 85 K is approximately six times lower than at 400 K. For niobium, QE variation was relatively small, measured as $1.6 \cdot 10^{-5} \pm 1.5 \cdot 10^{-6}$, indicating a variation of only ±9% across the entire temperature range. It is worth noting that the vacuum conditions during the copper test were nearly one order of magnitude poorer than those during the niobium test. This raises the possibility that increased surface contamination of the copper sample may have contributed significantly to the observed QE discrepancy between copper and niobium.

Table 1: Calculated QE for Cu crystals.

| Crystal orientation | Unit | 100[a] | 110 | 111 | 112 |
|---|---|---|---|---|---|
| WF $E_c = 0$ | eV | 4.59 | 4.48 | 4.94 | 4.53 |
| QE at 300 K $E_c = 1.5$ MV/m | $10^{-4}$ | 1.1 | 2.2 | 0.077 | 1.7 |
| QE at 2 K $E_c = 35$ MV/m | $10^{-4}$ | 3.0 | 4.7 | 0.16 | 3.9 |

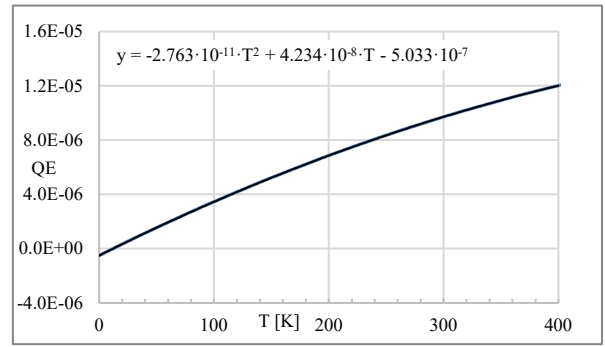

Figure 8: Fitting parabolic curve for the copper data and its mathematical formula. For T ≤ 12 K QE ≤ 0. *Courtesy A. Kara and J. R. Harris.*

**b.  Cathode-plug made of niobium**

Table 2 presents the calculated QE values for niobium crystals. As with copper, QE was calculated for both ambient temperature and 2 K using the same theoretical model. The optical data sources were consistent with those used for copper, with additional data for niobium taken from [23], and cryogenic optical properties sourced from [24, 25].

The QE calculations were based on photon energy of 4.82 eV and effective work functions $WF_{eff}$, corresponding to values $E_c = 1.5$ MV/m and $E_c = 35$ MV/m for 300 K and 2 K respectively.

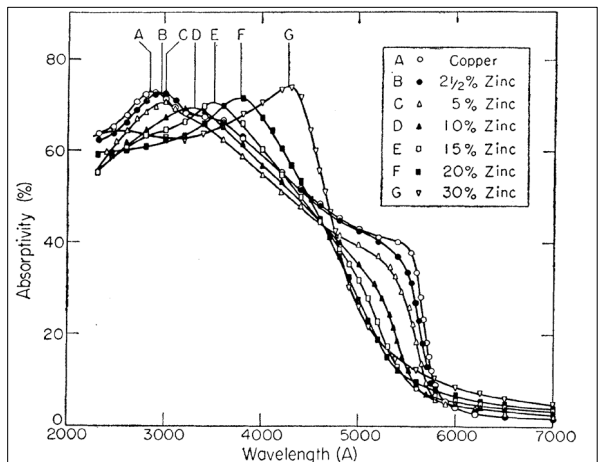

Figure 7: Absorptivity vs λ for polycrystalline copper and *α*-Brasses at 4.2 K. *Copy from [18].*

---

[a] For Cu 100-crystal WF=5.1eV is listed in several publications, e.g. [15, 16, 27]. The value in Table 1 is listed for example in [9, 17].

Table 2: Calculated QE for Nb crystals.

| Crystal orientation | Unit | 001 | 113 | 116 | 310 | 110 | 112 |
|---|---|---|---|---|---|---|---|
| WF $E_c = 0$ | eV | 4.02 | 4.29 | 3.95 | 4.18 | 4.87 | 4.63 |
| QE at 300 K $E_c$=1.5MV/m | $10^{-4}$ | 10 | 4.7 | 12 | 6.7 | 0.00002 | 0.79 |
| QE at 2 K $E_c$=35MV/m | $10^{-4}$ | 12.8 | 6.9 | 14.7 | 9.7 | 0.37 | 2.1 |

A QE of $5 \cdot 10^{-5}$ at 300 K for Nb polycrystal was experimentally measured at BNL in 2003 using 266 nm laser and $E_c$ of 1.5 MV/m [5]. Using the same parameters in our formula, the calculated QE was $2.1 \cdot 10^{-5}$ in reasonable agreement with the measured value. At 2 K, for λ of 248 nm and $E_c$ = 6 MV/m, the same authors reported a QE of only $3 \cdot 10^{-5}$, which was significantly lower than anticipated. Surface contamination was suggested as a possible explanation. Our theoretical model, however, predicts a QE of $6.4 \cdot 10^{-4}$, for polycrystalline niobium under similar conditions—more than an order of magnitude higher.

Figure 9 illustrates the dependence of calculated QE on the $WF_{eff}$ for both niobium and copper crystals and polycrystals, at 2 K and 300 K, using 257 nm laser and the relevant $E_c$ values for each temperature.

For polycrystalline metals (unfilled markers in Figure 9), QE values were derived using the arithmetic mean of the effective work functions of the corresponding crystals. This approach assumes that all crystallographic orientations are equally represented in the polycrystalline structure. However, in practice, this may not be the case due to the specific metallurgical processes used during fabrication, and subsequent surface treatments such as annealing or polishing. These factors can lead to preferential grain orientations and variations in surface condition, which in turn affect photoemission properties. This consideration becomes particularly relevant for cathodes used in ultrafast electron diffraction (UED) facilities, where the laser spot diameter is typically limited to several tens of microns in order to generate electron bunches with femtosecond-scale durations. In such cases, the QE may be influenced by the dominant grain orientation within the illuminated region, highlighting the need for careful material selection and surface preparation.

### c. Criteria for the choice of cathode

Any improvement in QE reduces the required laser power absorbed by the cathode, thereby enhancing thermal stability and allowing the use of lower-power, less demanding laser systems.

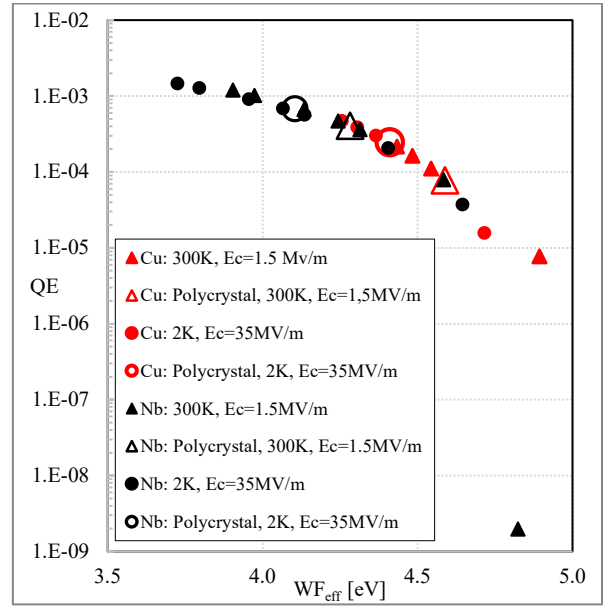

Figure 9: Calculated QEs for crystals and polycrystals of Nb and Cu at 300 K and 2 K and $E_c$ =1.5 MV/m and 35 MV/m, for λ = 257 nm.

This criterion is of particular importance for FEL and THz facilities—especially those operating, or designed to operate, in continuous wave (cw) mode with bunch charges of 50–250 pC and repetition rates up to 1 MHz. In such facilities, selecting crystal orientations with inherently higher QE than commonly used polycrystalline materials could serve as a primary optimisation strategy. For copper, the (1,1,0) orientation, and for niobium, the (0,0,1) orientations, appear to be favourable choices. In contrast, the QE requirement is less critical for UED facilities, which typically operate with bunch charges three orders of magnitude lower than those in FEL and THz systems.

Thermal stabilisation of the metallic cathode in superconducting RF (SRF) injectors must also be considered a design and operational constraint.

As an example, consider an electron beam with bunches of 100 pC at a 100 kHz repetition rate. For cathodes made from the aforementioned crystals and irradiated with 4.82 eV photons, the required laser power would be approximately 0.075 W for niobium and 0.21 W for copper. These values assume a conservative estimate of 50% of the theoretical QE values listed in Tables 1 and 2. Based on the measured absorptivity of Nb [25] and Cu [Fig. 7], the reflectivity of both metals at cryogenic temperatures for photons with an energy of 4.82 eV is approximately 55% for Nb and 37% for Cu. Considering both RF power dissipation at $E_c$=35 MV/m and laser light absorption, the total power dissipation would be

~ 34 mW for the niobium cathode and ~185 mW for the cathode made from copper with RRR = 100. Under these conditions, in the steady state, the cathode temperatures would increase to ~ 2.32 K and ~ 2.97 K, respectively. In both scenarios, the SRF gun cavity remains thermally stable with a substantial safety margin.

An alternative method for increasing QE is to operate the gun cavity at a higher accelerating gradient, thereby exposing the cathode to a stronger electric field $E_c$, which reduces the effective work function $WF_{eff}$ via the Schottky effect. This approach is viable provided that higher gradients do not induce dark current generation. Figures 3 and 4 show that, for the current cavity geometry, high-performance prototypes have achieved gradients up to 56 MV/m using either Cu or Nb cathodes. However, strong radiation was observed at these levels, attributed to electron field emission, which begins at approximately 40–45 MV/m. In the following section, we will discuss a proposed modification to the cavity geometry in the region near the cathode. This modification is designed to significantly reduce the electric field on the cavity wall, thereby lowering the probability of dark current while preserving high accelerating gradients at the cathode surface.

### III. Proposed new shape of the cavity

Figure 10 presents the current RF profile of the DESY gun cavity. All RF modelling for the cavities was performed using a 2D finite element method (FEM) code, as described in [28]. Since the modelled section of the cavity is cylindrically symmetric, the simulations were conducted in cylindrical coordinates (r, z, ϕ).

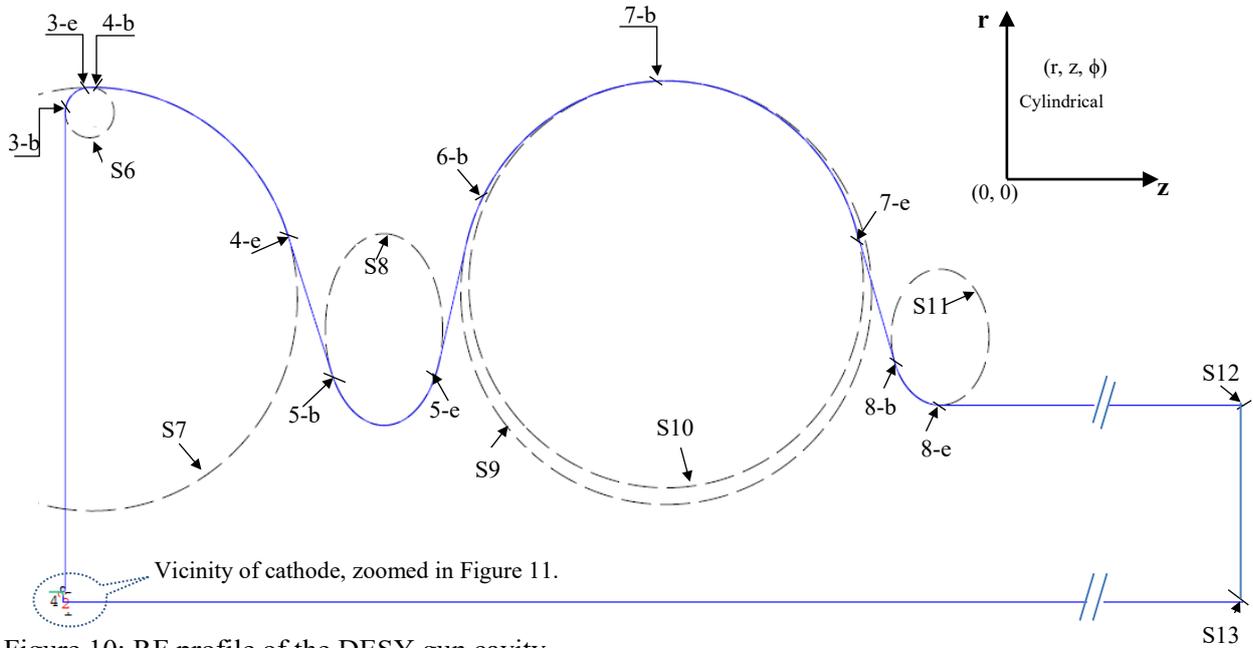

Figure 10: RF profile of the DESY gun cavity

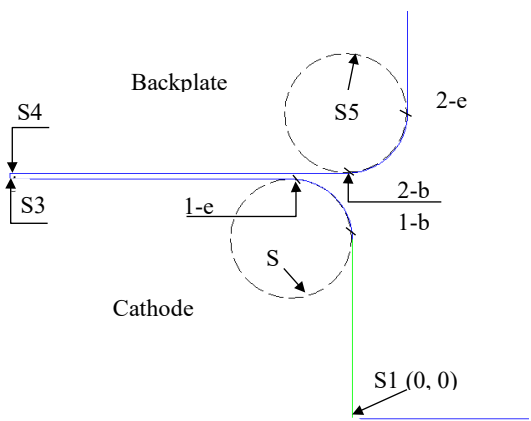

Figure 11: Zoomed-in circular RF profile of the DESY gun cavity in the vicinity of the cathode.

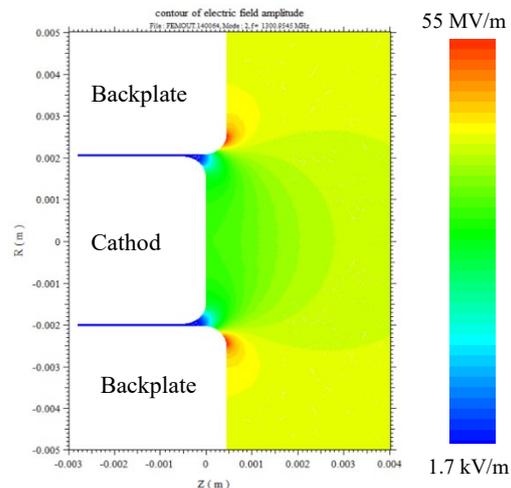

Figure 12: Electric field near the cathode at a peak on-axis field of 40 MV/m.

Figure 11 shows a zoomed-in view of the cathode region, where the cathode is retracted by 450 µm into the cathode channel. The retraction of cathodes with a 4 mm diameter was initially investigated in 2016 as a potential method for compensating emittance growth. A retraction depth of 450 µm was identified as optimal and has been implemented in all prototypes since 2017, starting with prototypes 16G03 and 16G04. More detailed studies conducted in 2023 reaffirmed the validity of this optimisation [29]. The coordinates (r, z) of key points (labelled 1-b to 8-e) and segments (S1 to S13) shown in Figures 10 and 11 are listed in Table 3. The segments represent points that define the start of straight lines, circles, and ellipses forming the cavity's geometry.

Figure 12 displays the electric field near the cathode when the peak electric field on the axis of both cells is 40 MV/m. Under this condition, the peak electric field on the cavity wall reaches 55 MV/m. The field enhancement is 38%.

Figure 13 illustrates the modified cavity geometry, which differs from the current design by replacing the circular edge of the cathode channel with an elliptical contour. The coordinates for the points marked in Figures 10 and 13 are provided in Table 4. This geometric alteration leads to a significant reduction in the peak electric field on the cavity wall, as shown in Figure 14. The resulting field enhancement is reduced to just 7.5%.

This improvement is expected to positively impact the performance of injectors based on the DESY gun cavity design. The ability to operate at higher gradients enables the production of electron bunches with charges exceeding 250 pC and energies greater than 4 MeV, while maintaining low emittance. In facilities operating with very low charge—on the order of ~100 fC—minimising dark current is essential to prevent background noise, particularly in experiments that require few-fC-level bunches.

Further studies of the proposed cavity shape confirmed the absence of multipacting in the cathode region for both copper and niobium cathodes across the gradient range of 5 MV/m to 55 MV/m. In addition, beam dynamics simulations for injectors based on both the circular and elliptical cavity geometries—operating at $E_c$ = 50 MV/m and followed by a standard E-XFEL cryomodule containing eight TESLA cavities at 32 MV/m—indicate that both configurations produce high-quality bunches. However, the elliptical design offers a modest performance advantage.

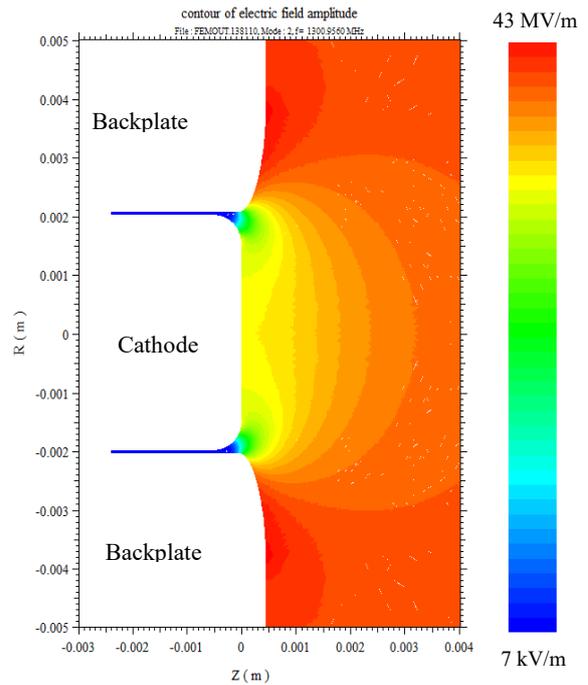

Figure 14: Electric field near the cathode at 40 MV/m peak on-axis field for the new elliptical shape.

The resonant frequency of the accelerating mode for the modified shape differs from that of the current design by less than 2 kHz. The RF power dissipated at the cathode increases by approximately 13% in the modified geometry, a manageable trade-off given the improved field distribution and dark current suppression.

For the next two prototypes, 16G11 and 16G12—currently in the production phase—the modified elliptical geometry will be implemented, provided that fine-grain niobium for the backplate becomes available in time. Upcoming cryogenic tests using both niobium and copper cathodes will validate whether the predictions made in our modelling studies for the new shape hold true in practical operation.

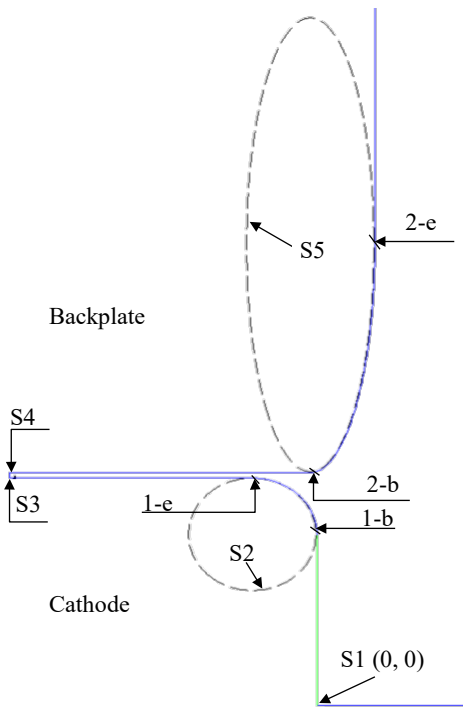

Figure 13: Zoomed-in new elliptical RF profile of the DESY gun cavity in the vicinity of the cathode.

## IV. Summary

The metallic cathodes presented—manufactured from single-crystal copper or single-crystal niobium—appear to be highly promising candidates for application in the DESY SRF photoinjector cavity. Theoretical QE is higher for niobium crystals compared with copper, primarily due to niobium's lower work functions across various crystal orientations

The principal advantage of pure copper lies in its substantially higher thermal conductivity at cryogenic temperatures. Nonetheless, earlier static thermal studies carried out at DESY demonstrated that even with a power dissipation of 1.2 W in the niobium cathode, the temperature only rose to approximately 4.1 K. This is because niobium's thermal conductivity increases sharply with temperature; at 4.1 K, it is already five times greater than at 2 K, thus effectively supporting thermal stabilisation despite lower baseline conductivity.

In the second part of this note, we revisited the comparison between the existing gun cavity geometry and a slightly modified version. The modified design features an elliptical contour near the cathode channel, which reduces the surface electric field enhancement in this region from 38% (in the current geometry) to just 7.5%. This reduction in peak surface field is of practical importance. The adoption of the new geometry would allow for stable operation at higher accelerating gradients, reduce or even eliminate dark current generation, and enable further exploitation of the Schottky effect by increasing the cathode's electric field. This, in turn, would effectively lower the work function of the implemented crystalline cathodes, improving QE without compromising cavity performance.

## Acknowledgments

The authors wish to express their sincere gratitude to the DESY team—M. Schmoekel, B. van der Horst, J. Iversen, and J. H. Thie—for their ongoing support and invaluable assistance throughout the course of this project.

We would also like to extend our appreciation to those who contributed significantly in the early stages of the programme. In particular, we acknowledge the support and collaborative expertise of the team from BNL, including T. Rao, J. Smedley (now at SLAC), and Ilan Ben-Zvi. Special thanks are due to P. Kneisel from TJNAF for his guidance in SRF technology and his participation in many crucial experiments, and to J. Lorkiewicz of the National Centre for Nuclear Research (NCBJ) for providing numerous lead cathodes.

Table 3: Coordinates of the market points and segments for circular shape. See Figures 10 and 11.

| Segment | Point | Segment type | Flag for the input | r [m] | z [m] |
|---|---|---|---|---|---|
| S1 | 1 | Beginning of line | 0 | 0 | 0 |
| S2 | 1-b | Circle #1 | 2 | 0.0015 | 0 |
|  | 1-e | Circle #1 | 0 | 0.002 | -0.0005 |
| S3 | 3 | Beginning of line | 0 | 0.002 | -0.0028 |
| S4 | 4 | Beginning of line | 0 | 0.00205 | -0.0028 |
| S5 | 2-b | Circle #2 | 2 | 0.00205 | -0.00005 |
|  | 2-e | Circle #2 | 0 | 0.00255 | 0.00045 |
| S6 | 3-b | Circle #3 | 2 | 0.0938 | 0.00045 |
|  | 3-e | Circle #3 | 0 | 0.1008 | 0.00745 |
| S7 | 4-b | Ellipse #4 | 2 | 0.1008 | 0.00790 |
|  | 4-e | Ellipse #4 | 0 | 0.070079 | 0.048850 |
| S8 | 5-b | Ellipse #5 | 2 | 0.047193 | 0.054397 |
|  | 5-e | Ellipse #5 | 0 | 0.0473371 | 0.076838 |
| S9 | 6-b | Circle #6 | 2 | 0.0709714 | 0.082428 |
| S10 | 7-b | Circle #7 | 2 | 0.103305 | 0.1233 |
|  | 7-e | Circle #7 | 0 | 0.0744981 | 0.161956 |
| S11 | 8-b | Ellipse #8 | 2 | 0.0474563 | 0.170024 |
|  | 8-e | Ellipse #8 | 0 | 0.039 | 0.1793 |
| S12 | 12 | Beginning of line | 0 | 0.039 | 0.3199* |
| S13 | 13 | End line | -2 | 0 | 0.3199* |
|  | $r_o$ centre | $z_o$ centre | Inclination Angle | Half axis $h_r$ | Half axis $h_z$ |
|  | [m] | [m] | [°] | [m] | [m] |
| Circle #1 | 0.0015 | -0.0005 | 0 | 0.0005 | 0.0005 |
| Circle #2 | 0.00255 | -0.00005 | 0 | 0.0005 | 0.0005 |
| Circle #3 | 0.0938 | 0.00745 | 0 | 0.007 | 0.007 |
| Ellipse #4 | 0.0613 | 0.0079 | 0 | 0.0395 | 0.042 |
| Ellipse #5 | 0.054 | 0.0656 | 0 | 0.019 | 0.012 |
| Circle #6 | 0.061305 | 0.1233 | 0 | 0.042 | 0.042 |
| Circle #7 | 0.062965 | 0.1233 | 0 | 0.04034 | 0.04034 |
| Ellipse #8 | 0.0525 | 0.1793 | 0 | 0.0135 | 0.01 |

Legend: 0-line, 1-circle, 2-ellipse, -2 end, n-b beginning of segment, n-e end of segment.
* The length of the beam tube can be adjusted, but should be longer than 100 mm.
The π-mode frequency for the evacuated cavity is 1301.00 GHz, and field flatness is 99%.

Table 4: Coordinates of the market points and segments for elliptical shape. See Figures 10 and 13.

| Segment | Point | Segment type | Flag for the input | r [m] | z [m] |
|---|---|---|---|---|---|
| S1 | 1 | Beginning of line | 0 | 0 | 0 |
| S2 | 1-b | Circle #1 | 2 | 0.0015 | 0 |
|  | 1-e | Circle #1 | 0 | 0.002 | -0.0005 |
| S3 | 3 | Beginning of line | 0 | 0.002 | -0.0028 |
| S4 | 4 | Beginning of line | 0 | 0.00205 | -0.0028 |
| S5 | 2-b | Ellipse #2 | 2 | 0.00205 | -0.00005 |
|  | 2-e | Ellipse #2 | 0 | 0.00405 | 0.00045 |
| S6 | 3-b | Circle #3 | 2 | 0.0938 | 0.00045 |
|  | 3-e | Circle #3 | 0 | 0.1008 | 0.00745 |
| S7 | 4-b | Ellipse #4 | 2 | 0.1008 | 0.00790 |
|  | 4-e | Ellipse #4 | 0 | 0.070079 | 0.048850 |
| S8 | 5-b | Ellipse #5 | 2 | 0.047193 | 0.054397 |
|  | 5-e | Ellipse #5 | 0 | 0.0473371 | 0.076838 |
| S9 | 6-b | Circle #6 | 2 | 0.0709714 | 0.082428 |
| S10 | 7-b | Circle #7 | 2 | 0.103305 | 0.1233 |
|  | 7-e | Circle #7 | 0 | 0.0744981 | 0.161956 |
| S11 | 8-b | Ellipse #8 | 2 | 0.0474563 | 0.170024 |
|  | 8-e | Ellipse #8 | 0 | 0.039 | 0.1793 |
| S12 | 12 | Beginning of line | 0 | 0.039 | 0.3199* |
| S13 | 13 | End line | -2 | 0 | 0.3199* |
|  | $r_o$ centre | $z_o$ centre | Inclination Angle | Half axis $h_r$ | Half axis $h_z$ |
|  | [m] | [m] | [°] | [m] | [m] |
| Circle #1 | 0.0015 | -0.0005 | 0 | 0.0005 | 0.0005 |
| Circle #2 | 0.00255 | -0.00005 | 0 | 0.0005 | 0.002 |
| Circle #3 | 0.0938 | 0.00745 | 0 | 0.007 | 0.007 |
| Ellipse #4 | 0.0613 | 0.0079 | 0 | 0.0395 | 0.042 |
| Ellipse #5 | 0.054 | 0.0656 | 0 | 0.019 | 0.012 |
| Circle #6 | 0.061305 | 0.1233 | 0 | 0.042 | 0.042 |
| Circle #7 | 0.062965 | 0.1233 | 0 | 0.04034 | 0.04034 |
| Ellipse #8 | 0.0525 | 0.1793 | 0 | 0.0135 | 0.01 |

Legend: 0-line, 1-circle, 2-ellipse, -2 end, n-b beginning of segment, n-e end of segment.
* The length of the beam tube can be adjusted, but should be longer than 100 mm.
The π-mode frequency for the evacuated cavity is 1301.00 GHz, and field flatness is 99%.

# References

[1] J. Sekutowicz et al., "Nb-Pb Superconducting RF-gun", TESLA-FEL Report 2005-09, DESY, Hamburg, 2005. https://flash.desy.de/sites2009/site_vuvfel/content/e403/e1642/e713/e714/infoboxContent880/fel2005-09.pdf

[2] J. Sekutowicz, G. Ciovati, P. Kneisel, G. Wu, A. Brinkmann, R. Parodi, W. Hartung, S. Zheng, "Cavities for JLab's 12 GeV Upgrade", Proc. PAC03, Portland, 2003. https://accelconf.web.cern.ch/p03/PAPERS/TPAB085.PDF

[3] J. Smedley, T. Srinivasan-Rao, J. Warren, R. S. Lefferts, A. R. Lipski, J. Sekutowicz, "Photoemission Properties of Lead", Proc. of the 9th EPAC04, Lucerne, Swiss, 2004. https://accelconf.web.cern.ch/e04/PAPERS/TUPKF080.PDF

[4] T. Srinivasan-Rao et al., "Design, Construction and Status of All Niobium Superconducting Photoinjector at BNL", Proc. PAC2003, Oregon, USA, 2003. https://accelconf.web.cern.ch/p03/PAPERS/MOPB010.PDF

[5] Q. Zhao, T. Srinivasan-Rao, M. Cole, "Tests of niobium cathode for the superconducting radio frequency gun", Proc. PAC2003, Oregon, USA, 2003. https://accelconf.web.cern.ch/p03/PAPERS/WPAB008.PDF

[6] T. Rao, I. Ben-Zvi, A. Burrill, H. Hahn, D. Kayran, Y. Zhao, P. Kneisel, M. Cole, "Photo emission Studies on BNL/AES/Jlab all Niobium Super-conducting RF Injector", Proc. PAC2005, Knoxville, Tennessee, 2005. https://accelconf.web.cern.ch/p05/PAPERS/WPAP038.PDF

[7] P. Davis et al., "Quantum Efficiency Measurements of a Copper Photocathode in an RF Electron Gun", Proc. PAC1993, Washington, https://apps.dtic.mil/sti/tr/pdf/ADA280385.pdf

[8] X.J. Wang et al., "Experimental characterization of the high-brightness electron photoinjector", NIM in PR A. Volume 375, Issues 1–3, 11 June 1996, Pages 82-86. https://www.sciencedirect.com/science/article/abs/pii/0168900296000393

[9] D. Palmer, S. Anderson, J. B. Rosenzweig, "Single crystal cooper photo-cathode in the BNL/SLAC/UCLA 1.6-cell rf gun", Proc. of the 2nd ICFA AA Workshop, Los Angeles, 1999. https://www.worldscientific.com/doi/abs/10.1142/9789812792181_0028

[10] A. Brachmann et al., "LCLS RF gun copper cathode performance", Proc. of the 2nd IPAC, San Sebastian, Spain, 2011, pp. 3200-3202. https://accelconf.web.cern.ch/IPAC2011/papers/thpc134.pdf

[11] E. Vogel et al., "High gradients at SRF photo-injector cavities with low RRR copper cathode plug screwed to the cavity backplate", 2023. http://arxiv.org/abs/2310.02974

[12] Simon, N. J.," Properties of Copper and Copper Alloys at Cryogenic Temperatures", NIST Monograph, Publication 177, 1992, Section 7, p. 19. https://www.govinfo.gov/app/details/GOVPUB-C13-50e4c1dff3302ac3cc5de7c29c768576

[13] D. H. Dowell and J. F. Schmerge, "Quantum Efficiency and Thermal Emittance of Metal Photocathodes", PR ST - Accelerators and Beams 12, 074201, 2009. https://journals.aps.org/prab/abstract/10.1103/PhysRevSTAB.12.074201

[14] J. Wang and S.Q. Wang, "Surface energy and work function of fcc and bcc crystals: Density functional study", Surface Science, December 2014. https://www.sciencedirect.com/science/article/abs/pii/S0039602814002441?via%3Dihub

[15] F. Le Pimpec et al.," Quantum efficiency of technical metal photocathodes under laser irradiation of various wavelength", 5 Oct 2012. http://arxiv.org/abs/1202.0152v2

[16] W. M. Haynes et al., "CRC Handbook of Chemistry and Physics", CRC Press Taylor & Francis Group, 97's Edition 2016-2017. https://doi.org/10.1201/9781315380476

[17] H. B. Michaelson, "The Work Function of the Elements and its Periodicity, J. Appl. Phys. 48, 4729–4733, 1977. https://doi.org/10.1063/1.323539

[18] M. Biondi J. Rayne, "Band Structure of Noble Metal Alloys: Optical Absorption in $\alpha$-Brasses at 4.2°K", Phys. Rev. 115, 1522, September 1959. https://journals.aps.org/pr/pdf/10.1103/PhysRev.115.1522

[19] M. A. Biondi, "Optical Absorption of Copper and Silver at 4.2 K", Physical Review Volume 102, Number 4, May 15, 1956. https://journals.aps.org/pr/abstract/10.1103/PhysRev.102.964

[20] M. A. Biondi, "Optical and Infrared Absorption of Copper at 4.2°K", Phys. Rev. 96, 534 – Published 15 October, 1954. https://journals.aps.org/pr/abstract/10.1103/PhysRev.96.534

[21] J. R. Harris et al., "Temperature Dependence of Photo-emission from Copper and Niobium", Proc. of PAC2013, Pasadena, CA USA. https://accelconf.web.cern.ch/pac2013/papers/thpac19.pdf

[22] A. Kara, "Quantum Efficiency as a Function of Temperature in Metal Photocathodes", Thesis, NPS, June 2013. https://apps.dtic.mil/sti/tr/pdf/ADA585521.pdf

[23] J. Smedley, T. Rao, Q. Zhao, "Photoemission studies on niobium for superconducting photoinjectors", J. Appl. Phys. 98, 043111, 2005. https://doi.org/10.1063/1.2008389

[24] H. Padamsee, J. Knobloch, T. Hays, "RF super-conductivity for accelerators", John Wiley & Sons, INC., 1998, 2nd Edition 2008. https://www.wiley.com/en-us/RF+Superconductivity+for+Accelerators%2C+2nd+Edition-p-9783527408429

[25] J. Weaver, D. Lynch, C. Olson, "Optical Properties of Niobium from 0.1 to 36.4 eV", Phys. Rev. B 7, 4311, 15 May, 1973.
https://doi.org/10.1103/PhysRevB.7.4311

[26] P. B. Johnson and R. W. Christy, "Optical Constants of the Noble Metals", Physical Review B Volume 6, Number 12 December 1972,
https://journals.aps.org/prb/pdf/10.1103/PhysRevB.6.4370

[27] G. Gaertner, W. Knapp, R. G. Forbes, "Modern Developments in Vacuum Electron Sources", Topics in Applied Physics, Volume 135, Springer, 2020.
https://link.springer.com/content/pdf/10.1007/978-3-030-47291-7.pdf

[28] J. Sekutowicz, "2D FEM Code with Third Order Approximation for RF Cavity Computation", Proc. of the International Linac Conference, Tsukuba, Japan,1994.
https://accelconf.web.cern.ch/l94/papers/mo-88.pdf

[29] E. Gjonaj, D. Bazyl, "Beam Dynamics Study of a cw L-Band SRF Gun for the High Duty Cycle EUXFEL", Proc. IPAC14, Venice, Italy, 2023.
https://accelconf.web.cern.ch/ipac2023/pdf/WEPA054.pdf